\documentstyle [twocolumn,epsf]{mn}
\newcommand{\mincir}{\raise
-3.truept\hbox{\rlap{\hbox{$\sim$}}\raise4.truept\hbox{$<$}\ }}
\newcommand{\magcir}{\raise
-3.truept\hbox{\rlap{\hbox{$\sim$}}\raise4.truept\hbox{$>$}\ }}
\newcommand{\minmag}{\raise
-3.truept\hbox{\rlap{\hbox{$<$}}\raise5.truept\hbox{$<$}\ }}
\newcommand{\be}{\begin{equation}}
\newcommand{\ee}{\end{equation}}
 \newcommand{\ba}{\begin{eqnarray}}
\newcommand{\ea}{\end{eqnarray}}
\newcommand{\brr}{\begin{array}}
\newcommand{\err}{\end{array}}
\newcommand{\bc}{\begin{center}}
\newcommand{\ec}{\end{center}}

\renewcommand{\baselinestretch}{1.0}

\title[Cosmological implications and structure formation 
from a time varying vacuum]
{Cosmological implications and structure formation from a time varying vacuum}
\author[Spyros Basilakos]{Spyros Basilakos \\
\vspace{0.1cm}
Academy of Athens, Research Center for Astronomy \& Applied
  Mathematics, Soranou Efessiou 4, 11-527, Athens, Greece\\
}

\begin{document}

\maketitle

\begin{abstract}
We study the dynamics of the FLRW flat cosmological models 
in which the vacuum energy varies with time, $\Lambda(t)$.
In this model we find that the main cosmological functions such as the scale 
factor of the universe and the Hubble flow are defined in 
terms of exponential functions. 
Applying a joint likelihood analysis 
of the recent supernovae type Ia data, the Cosmic Microwave Background
shift parameter  
and the Baryonic Acoustic Oscillations traced by the Sloan Digital 
Sky Survey (SDSS) galaxies, 
we place tight constraints on the main cosmological parameters 
of the $\Lambda(t)$ scenario. 
Also, we compare the $\Lambda(t)$ model with the traditional 
$\Lambda$ cosmology and we find that the former model provides a Hubble
expansion which compares well with that of the $\Lambda$ cosmology.
However, the $\Lambda(t)$ scenario predicts stronger small scale
dynamics, which implies a faster growth rate of perturbations with respect 
to the usual $\Lambda$-cosmology, despite the fact that they
share the same equation of state parameter.
In this framework, we find that galaxy clusters in the $\Lambda(t)$ model 
appear to form earlier than in the $\Lambda$ model. 

\end{abstract}

\vspace{1.0cm}

\section{Introduction}
The detailed analysis of the available high quality cosmological 
observations 
(Riess et al. 1998; 
Perlmutter et al. 1999; Efstathiou et al. 2002;
Basilakos \& Plionis 2005; Tegmark et al. 2006; 
Davis et al. 2007, Kowalski et al. 2008; Komatsu et al. 2009) 
have converged during the last decade towards a cosmic expansion
history that involves a spatial flat geometry and 
a recent accelerating expansion of the
universe. This expansion has been attributed to an energy component
(dark energy) with negative pressure which dominates the universe at
late times and causes the observed accelerating expansion. The
simplest type of dark energy corresponds to the cosmological
constant (see for review Peebles \& Ratra 2003). 
The nature of the dark energy is still a mystery and indeed it is 
one of the most fundamental current problems in physics and cosmology. 

In the literature, there are many theoretical speculations 
regarding the physics of the above exotic dark energy.
The simplest approach is to consider a  
real scalar field $\phi$ which rolls down the potential energy 
$V(\phi)$ and therefore it could mimic the dark energy 
(Ratra \& Peebles 1988; Weinberg 1989; Turner \& White 1997;
Caldwell, Dave \& Steinhardt 1998; Padmanabhan 2003). 
Alternatively, Ozer \& Taha (1987) proposed a different scenario in which 
a time varying $\Lambda$ parameter could be a possible 
candidate for the dark energy (see also Bertolami 1986; Freese et al. 1987; 
Peebles \& Ratra 1988; 
Carvalho, Lima \& Waga 1992; Overduin \& Cooperstock 1998;
Bertolami \& Martins 2000; Alcaniz \& Maia 2003; Opher \& Pellison 2004; 
Bauer 2005; Barrow \& Clifton 2006;
Montenegro \& Carneiro 2007 and references therein). In this 
cosmological model the dark energy equation
of state parameter $w\equiv P_{DE}/\rho_{DE}$,
is strictly equal to -1, but the vacuum energy density (or $\Lambda$) 
varies with time. It is interesting to mention here
that the renormalization group (RG) 
in quantum field theory (Shapiro \& Sol\'a 2000; Babi\'c et al. 2002) 
provides a time varying vacuum, in which the $\Lambda$ 
component evolves as $\sim H^{2}(t)$ [see Grande, Sol\'a \& Stefancic 2006] 
, where $H$ is the Hubble parameter. 
On the other hand, based on the holographic principle 
(Bousso 2002; Padmanabhan 2005) one can prove that 
$\Lambda \sim H^{4}$.

However, in the $\Lambda(t)$ cosmological model there is a coupling 
between the time-dependent vacuum and matter (Carneiro et al. 2008). 
In particular, using the combination of the conservation of the total energy
with the variation of the vacuum energy one can prove that 
the $\Lambda(t)$ model provides either a process of a particle production 
or the mass of the dark matter particles increases.
The latter general properties can be explained within the framework 
of the interacting dark energy models (Alcaniz \& Lima 2005 
and references therein).
We would like to stress here that 
most of the recent papers in dark energy studies are based 
on the assumption that
the dark energy evolves independently of the dark matter. Of course, 
the unknown nature of both dark matter and dark energy 
implies that at the moment 
we can not exclude the possibility of 
interactions in the dark sector. The confirmation of such a 
possibility would be of paramount importance 
because interactions between dark matter and 
dark energy could provide possible solutions 
to the cosmological coincidence problem. In general, 
several papers have been published in this area 
(eg., Zimdahl, Pav\'on, Chimento 2001; 
Amendola et al. 2003; Cai \& Wang 2005; Binder \& Kremer 2006; Das, 
Corasaniti, \& Khoury 2006; Olivares, 
Atrio-Barandela \& Pav\'on 2008 and references therein)
proposing that the dark energy and dark matter could be coupled.

The aim of the present work is to investigate the observational consequences 
of the overall dynamics by using the $\Lambda(t)$ cosmological model.
Due to the absence of a physically well-motivated 
functional form for the $\Lambda(t)$ parameter, we consider 
a power series form in $H$ up to a second order. Doing so, 
we include the effects of the de-Sitter spacetime.
The plan of the paper is as follows. 
The basic theoretical elements of the problem are 
presented in section 2 by solving analytically [for a spatially flat 
Friedmann-Lemaitre-Robertson-Walker (FLRW) geometry]
the basic cosmological equations. 
In section 3 we place constraints on the main parameters of our model by
performing a joint likelihood analysis utilizing the Union08 SNIa data 
(Kowalski et al. 2008), the shift parameter of the Cosmic Microwave Background 
(Komatsu et al. 2009) and the observed Baryonic Acoustic Oscillations 
(BAOs; Eisenstein et al. 2005;  Padmanabhan, et al. 2007).
Section 4 outlines the comparison 
between the time varying vacuum model with the traditional $\Lambda$ cosmology.
Also, in section 4 we solve analytically the time evolution equation
of the mass density contrast for the $\Lambda(t)$ model while in section 5
we present theoretical predictions regarding the formation of 
the galaxy clusters. In section 6 we
draw our conclusions. Finally, in the appendix we have treated
analytically, the basic cosmological equations 
considering that the time varying $\Lambda(t)$ parameter can be expressed 
with the aid of a power series expansion in $H$ up to a third order.
Note, that throughout the paper we use $H_{0}=70.5$Km/sec/Mpc
(Freedman et al. 2001; Komatsu et al. 2009).

\section{Cosmology with a time dependent vacuum}
In the framework, of a spatially flat 
Friedmann-Lemaitre-Robertson-Walker (FLRW) geometry the basic equations 
which governs the global dynamics of the universe are
\be
\rho_{m}+\rho_{\Lambda}=3H^{2}
\label{frie1} 
\ee
and 
\be 
\frac{d({\rho}_{m}+\rho_{\Lambda})}{dt}+\+3H(\rho_{m}+P_{m}+
\rho_{\Lambda}+P_{\Lambda})=0
\label{frie2} 
\ee
where $\rho_{m}$ and $\rho_{\Lambda}$ are the matter density and 
vacuum density respectively, while $P_{m}=0$ and $P_{\Lambda}$ is 
the corresponding vacuum pressure. 
Note, that for simplicity we use geometrical units 
($8\pi G=c\equiv 1$) in which $\rho_{\Lambda}=\Lambda$. 
In order to study the above system of differential equations we need to 
define explicitly the functional form of the $\Lambda$ component. 
Within the framework of the $\Lambda(t)$ model it is interesting to note 
that the equation of state takes the usual form of 
$P_{\Lambda}=-\rho_{\Lambda}(t)=-\Lambda(t)$ [Ozer \& Taha 1987; 
Peebles \& Ratra 1988].

On the other hand, introducing in the global dynamics the above idea
in a form of the time-dependent vacuum, 
it is possible to explain the physical properties of the 
dark energy.
Considering now  
eq.(\ref{frie2}), we have the following useful 
formula (see also Carneiro et al. 2008):
\be
\dot{\rho_{m}}+3H\rho_{m}=-\dot{\Lambda}
\label{frie33} 
\ee
and indeed, using eq.(\ref{frie1}), 
we obtain:
\be 
2\dot{H}+3H^{2}=\Lambda
\label{frie34} 
\ee
or
\be
\int_{+\infty}^{H}\frac{dy}{\Lambda-3y^{2}}=\int_{0}^{t} \frac{du}{2}=
\frac{t}{2}
\label{frie344} 
\ee
where the over-dot denotes derivatives with respect to time.
Of course, the traditional $\Lambda$ cosmology 
can be described by the above integration (eq.\ref{frie344})  
using a constant vacuum term $\Lambda=const$ 
(for more details see section 3.5). 

Now, from eq.(\ref{frie33}), it becomes evident that in this 
cosmological scenario there is a coupling between 
the time-dependent vacuum and matter. 
Actually, the idea for possible interactions in the dark sector is not really 
new in this kind of studies.
It has been shown that the coupling between dark matter and 
dark energy could provide possible solutions 
to the cosmological coincidence problem
(eg., Zimdahl, Pav\'on, Chimento 2001; 
Amendola et al. 2003; Cai \& Wang 2005; Binder \& Kremer 2006; Das, 
Corasaniti, \& Khoury 2006; Olivares, Atrio-Barandela \& Pav\'on 2008
and references therein). In this context, one of the most 
important issues and unknowns is the precise 
functional form of the equation of state parameter
$w(a)$ where ($a$ is the scale factor). 
The usual procedure is to derive a $w(a)$ approximate functional
form, by using a Taylor expansion around the present epoch  
(eg. Chevalier \& Polarski, 2001; Linder 2003), 
which then provides approximate 
solutions of the global density evolution.
However, the current approach is somewhat different in the sense that 
we do not ``design'' the equation of state parameter such
that to produce the desired (accelerated) cosmic evolution. 
Rather, we investigate whether a generalized vacuum component with 
$w(t)=-1$ and $P_{\Lambda}=-\rho_{\Lambda}(t)$ 
[Ozer \& Taha 1987; Peebles \& Ratra 1988] 
in the expanding Universe allows
for a late accelerated phase of the Universe 
and under which circumstances such
a solution provides a viable alternative 
to the dark energy.

Although, we do not have a fundamental theory to model the 
time-dependent $\Lambda(t)$ function, we can parametrize the latter
using a phenomenological approach. Indeed,  
in a series of recent papers, authors (see for example 
Ray, Mukhopadhyay \& Meng 2007; Sil \& Som 2008 and references therein)
have investigated the global dynamical properties of the universe
considering that the vacuum energy density decreases linearly 
either with the energy density or the square Hubble parameter.
Also, Wang \& Meng (2005), based on thermodynamics found that 
the vacuum energy density possibly decays as a power law. 
Alternatively, Carneiro et al. (2008) proposed 
a different pattern in which the vacuum term is proportional 
with the Hubble parameter, $\Lambda(a) \propto H(a)$. However, this model 
fails to fit the current CMB data (see also section 3.4). 
In this context, attempts to provide a theoretical explanation for 
the $\Lambda(t)$ have been presented in the 
literature (see Grande et al. 2006 and references therein). These 
authors found that a time dependent vacuum could
arise from the renormalization group (RG) in quantum field theory. The corresponding 
solution for a running $\Lambda(t)$
is found to be $\Lambda(t)\sim c_{1}H^{2}(t)$ [where $c_{1}$ is a constant] 
and it can mimic the quintessence or phantom behavior and transit smoothly between the two.
It is worth noting that at late enough times the above solution 
asymptotically reaches the de-Sitter regime $\Lambda \sim H^{2}$, as 
far as the global dynamics is concerned.

In this paper, we parametrize the functional form of $\Lambda(t)$ 
by applying a power series expansion in $H$ up to the second order
(see the appendix for a third order expansion which interestingly 
predict models with late acceleration but without 
initial singularities):
\be
\Lambda(t)=n_{1}H+n_{2}H^{2} \;\;.
\label{frie35} 
\ee
Obviously, eq.(\ref{frie35}) can be seen as a combination of the 
of the above ansatzs namely 
$H(t)$ (Carneiro et al. 2008) and $H^{2}(t)$ [quantum field theory; 
Grande et al. 2006] respectively. 
It is now routine to integrate eq.(\ref{frie344}) and obtain 
the Hubble function predicted by the current $\Lambda(t)$ model:
\be
H(t)=\frac{n_{1}}{\beta}\frac{{\rm e}^{n_{1}t/2}}{{\rm e}^{n_{1}t/2}-1} \;\;,
\label{frie4} 
\ee
where the range of $\beta(=3-n_{2})$ values for which the above integration
is valid is $\beta \in (0,+\infty)$ (or $n_{2}<3$). Of course, 
if we consider different patterns for the vacuum density then 
we can obtain different solutions for the Hubble parameter. 
Using now the definition of the Hubble parameter $H\equiv {\dot a}/a$, the 
scale factor of the universe $a(t)$, evolves with time as 
\be
a(t)=a_{1}\left({\rm e}^{n_{1}t/2}-1 \right)^{2/\beta} \;\;,
\label{frie44} 
\ee
where $a_{1}$ is the constant of integration. As expected, at late 
enough times the above solution reduces to the de-Sitter universe.
Note, that for $\beta \longrightarrow 3$ and 
at early times the $\Lambda(t)$ model tends to the 
Einstein de-Sitter case. Now from 
eqs.(\ref{frie4}, \ref{frie44}) we can easily write the 
corresponding Hubble flow
as a function of the scale factor
\be
H(a)=\frac{n_{1}}{\beta}\left[1+\left(\frac{a}{a_{1}}\right)^{-\beta/2} \right] \;\;.
\label{frie5} 
\ee
Evaluating eq.(\ref{frie5}) at the present time ($a\equiv1$) we obtain 
\be
n_{1}=\frac{\beta H_{0}}{1+a_{1}^{\beta/2}}
\label{frie6} 
\ee
where $H_{0}$ is the Hubble constant.
From eqs.(\ref{frie5}, \ref{frie6}), using the 
usual unit-less $\Omega$-parameterization, we have after some algebra
that:
\be
E(a)\equiv \frac{H(a)}{H_{0}}=\left(1-\Omega_{m}+\Omega_{m}a^{-\beta/2} \right)
\label{frie7} 
\ee
while the corresponding matter density parameter is: 
$\Omega_{m}(a)=\Omega_{m}a^{-\beta/2}/E(a)$. The normalized
scale factor of the universe becomes
\be
\label{frie8}
 a(t)=\left(\frac{\Omega_{m}}{1-\Omega_{m}}\right)^{2/\beta}
\left[{\rm e}^{\beta(1-\Omega_{m})H_{0}t/2}-1\right]^{2/\beta}
\ee
or
\be
\label{frie9}
t(a)=\frac{2}{\beta(1-\Omega_{m})}H^{-1}_{0} 
{\rm ln}\left[\frac{a^{\beta/2}E(a)}{\Omega_{m}} \right]
\ee
where $a_{1}^{\beta/2}=\Omega_{m}/(1-\Omega_{m})$.
It is interesting to point here that the current age of the 
universe [$a=1$, $E(1)=1$] is 
\be
t_{0}=\frac{2}{\beta}H^{-1}_{0} \frac{{\rm ln}\Omega_{m}}{\Omega_{m}-1} \;\;.
\ee
We now investigate the circumstances
under which an inflection point 
exists and therefore have an acceleration phase of the scale factor. 
This crucial period in the cosmic history corresponds to 
$\ddot{a}(t_{I})=0$. 
Differentiating twice eq.(\ref{frie8}), 
we then have:  
\be
a_{I}=\left[\frac{(\beta-2)\Omega_{m}}{2(1-\Omega_{m})}\right]^{2/\beta}
\;\; t_{I}=\frac{2}{\beta (1-\Omega_{m})}H^{-1}_{0}
{\rm ln}(\frac{\beta}{2}) 
\ee
which implies that the condition 
for which an inflection point is present in the evolution of 
the scale factor is $\beta>2$.

\begin{figure}
\mbox{\epsfxsize=8cm \epsffile{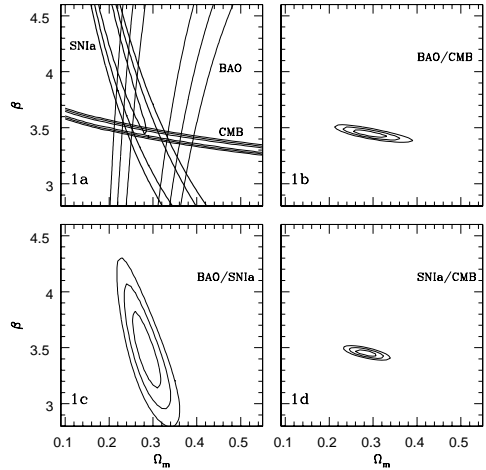}}
\caption{Likelihood contours in the $(\Omega_{m},\beta)$ plane.
The contours are plotted where 
$-2{\rm ln}({\cal L/L_{\rm max}})$ is equal to 2.32, 6.16 and 11.83,
corresponding to $1\sigma$, $2\sigma$ and $3\sigma$ confidence level. 
In Fig.1a we present the likelihood contours that 
correspond to the SNIa, CMB and BAOs observational data.
Finally, Figs. 1b, 1c and 1d show the statistical results for
different pairs.}
\end{figure}

\section{Cosmological Constraints}

\subsection{The likelihood from the CMB shift parameter}
A very accurate and deep geometrical probe of dark energy is the
angular scale of the sound horizon at the last scattering surface as
encoded in the location of the first peak of the
Cosmic Microwave Background (CMB) temperature perturbation spectrum. 
This probe is described by the so
called CMB shift parameter (cf. Bond, Efstathiou \& Tegmark 1997;
Trotta 2004; Nesseris \& Perivolaropoulos 2007) which is a normalized 
quantity and it is defined as
\be 
\label{shift}
R=\sqrt{\Omega_{\rm m}}\int_{a_{ls}}^1 \frac{da}{a^2
E(a)}=\sqrt{\Omega_{\rm m}}\int_0^{z_{ls}} \frac{dz}{E(z)}  \;\;.
\ee 
One of the merits of using the shift parameter in cosmological studies is 
that its dependence on the Hubble constant is negligible (for details 
see Melchiorri et al. 2003; Nesseris \& Perivolaropoulos 2007
and references therein). 
The shift parameter measured from the WMAP 5-years
data (Komatsu et al. 2009) is $R=1.71\pm 0.019$ at $z_{ls}=1090$  
[or $a_{ls}=(1+z_{ls})^{-1}\simeq 9.17\times 10^{-4}$]
and $E(z)\equiv H(z)/H_0$ is the normalized Hubble flow.
Therefore, the corresponding $\chi^{2}_{\rm cmb}$ function is simply written 
\begin{equation}
\chi^{2}_{\rm cmb}({\bf p})=\frac{[R({\bf p})-1.71]^{2}}{0.019^{2}}
\end{equation}
where ${\bf p}$ is a vector containing the cosmological
parameters that we want to fit.
Note, that we sample the unknown 
parameters as follows: $\Omega_{m} \in [0.1,1]$ 
and $\beta \in [2,5]$ in steps of 0.01. 
In Fig.1a we present the 1$\sigma$, 2$\sigma$ and $3\sigma$
confidence levels in the $(\Omega_{m},\beta)$ plane.
It is evident that the $\beta$ parameter is tightly
constrained ($\beta \simeq 3.58$) 
while the matter density parameter is not and all the values in the interval 
$0.1\le \Omega_{m} \le 1$ are acceptable (see Table 1).
However, following the WMAP 5-years results (Komatsu et al. 2009) 
of the full temperature perturbation spectrum $\Delta T/T$, we can use  
an additional constrain which is $\Omega_{m}h^{2}=0.1326\pm 0.0063$. 
Thus, for $h\simeq 0.71$ 
(Freedman et al. 2001; Komatsu et al. 2009) 
we find $0.24\le \Omega_{m} \le 0.29$ ($2\sigma$ limits).  

\subsection{The likelihood from the SNIa}
We now use the publicly available Union08 sample of 
307 supernovae of Kowalski et al. (2008) 
in order to constrain $\Omega_{m}$\footnote{http://supernova.lbl.gov/Union. 
This catalog includes the following components in the 
error budget of the distance moduli: (a) 
$\sigma_{\rm tot}$, (see Kowalski et al. 2008) obtained due to lensing, 
Milky way dust extinction and host galaxy peculiar velocities 
(b) the systematic error $\sigma_{\rm sys}$ and (c) the uncertainty which is related with 
the light-curve fitting. For the latter uncertainty,
Kowalski et al. take into account the stretch $s$ 
and color $c$ corrections via $\mu_{B}=m_{B}-M+\alpha(s-1)-bc$
for a specific cosmological model $(\Omega_{m},w)=(0.29,-0.97)$.
We would like to caution the reader that we do not
minimize $\chi^{2}$ over the parameters $\alpha$ and $b$, 
which implies that in the case of the $\Lambda(t)$ model we may not be
treating the third component of the error budget properly. 
However, we would like to stress that according to 
Kowalski et al. (2008) the corresponding constants 
$\alpha$ and $b$ are rather insensitive 
to the assumed cosmological parameters (see their section 5.1).
Thus, had we included the proper light-curve uncertainty in our fit we 
would have obtained a larger solution space (see figure 1a).}.
In this case, the likelihood function can be 
written as:
\begin{equation}
\label{chi22} 
\chi^{2}_{\rm SNIa}({\bf p})=\sum_{i=1}^{307} \left[ \frac{ {\cal \mu}^{\rm th}
(a_{i},{\bf p})-{\cal \mu}^{\rm obs}(a_{i}) }
{\sigma_{i}} \right]^{2} \;\;.
\end{equation}
where $a_{i}=(1+z_{i})^{-1}$ is the observed scale factor of
the universe, $z_{i}$ is the observed redshift, ${\cal \mu}$ is the 
distance modulus ${\cal \mu}=m-M=5{\rm log}d_{\rm L}+25$
and $d_{\rm L}(a,{\bf p})$ is the luminosity distance 
\begin{equation}
d_{\rm L}(a,{\bf p})=\frac{c}{H_{0}a} \int_{a}^{1} \frac{{\rm d}x}{x^{2}E(x)}
\end{equation}
where $c$ is the speed of light ($\equiv 1$ here).
Figure 1a also shows the 1$\sigma$, 2$\sigma$ and 3$\sigma$
confidence levels in the 
$(\Omega_{m},\beta)$ plane. Although, the $\beta$ parameter 
is not constrained by this analysis the matter density parameter 
has an upper limit of $\Omega_{m} \le 0.29$ 
within the $1\sigma$ uncertainty (see Table 1). 

\subsection{The likelihood from BAOs}
In this section, we utilize the so called Baryonic Acoustic Oscillations. BAOs 
are produced by pressure 
(acoustic) waves in the photon-baryon plasma in the 
early universe, generated by dark matter overdensities. 
Evidence of this excess has been found in the clustering 
properties of the luminous SDSS red-galaxies 
(Eisenstein et al. 2005;  Padmanabhan, et al. 2007) and it can provide a 
''standard ruler'' with which we can put constraints 
on the cosmological models. In particular,
we use the following estimator:
\begin{equation}
A({\bf p})=\frac{\sqrt{\Omega_{m}}}{[z^{2}_{s}E(a_{s})]^{1/3}}
\left[\int_{a_{s}}^{1} \frac{{\rm d}a}{a^{2}E(a)}
\right]^{2/3}
\end{equation}
measured from the SDSS data to be $A=0.469\pm 0.017$, 
where $z_{s}=0.35$ [or $a_{s}=(1+z_{s})^{-1}\simeq 0.75$].
In this case, the $\chi^{2}_{\rm BAO}$ function is given
\begin{equation}
\chi^{2}_{\rm BAO}({\bf p})=\frac{[A({\bf p})-0.469]^{2}}{0.017^{2}} \;\;.
\end{equation}
It is evident (see figure 1a), 
that the matter density parameter 
is constrained ($\Omega_{m} \simeq 0.28$) by this analysis, while 
the $\beta$ parameter is not (see also Table 1).

\begin{figure}
\mbox{\epsfxsize=8cm \epsffile{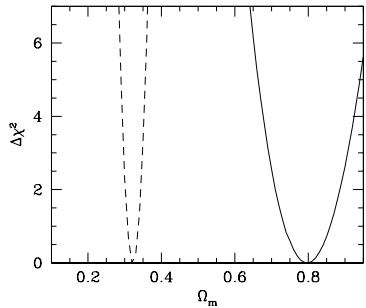}}
\caption{The variance $\Delta \chi^{2}=\chi^{2}-\chi^{2}_{min}$ 
around the best fit $\Omega_{m}$ value when we marginalize 
over $\beta=3$ [$\Lambda(t)\propto H(t)$]. The solid and the 
dashed line corresponds to the CMB and SNIa/BAO  
likelihoods respectively.} 
\end{figure}

\subsection{The joint likelihoods}
We can combine the above probes by using a joint likelihood analysis:
$${\cal L}_{tot}({\bf p})=
{\cal L}_{\rm BAO} \times {\cal L}_{\rm cmb}\times {\cal L}_{\rm SNIa} $$
or
$$\chi^{2}_{tot}({\bf p})=\chi^{2}_{\rm BAO}+\chi^{2}_{\rm cmb}+\chi^{2}_{\rm SNIa}$$ 
in order to put even further constraints on the parameter space used. 
Note, that we define the likelihood estimator\footnote{Likelihoods
are normalized to their maximum values. Note that the errors of the fitted 
parameters represent $1\sigma$ uncertainties.} as:
${\cal L}_{j}\propto {\rm exp}[-\chi^{2}_{j}/2]$.
The resulting best fit parameters, are presented in the second two rows of
Table 1. 
The overall likelihood function 
peaks at $\Omega_{m}=0.29^{+0.01}_{-0.02}$,
$\beta=3.44 \pm 0.02$ and
the corresponding $\chi_{tot}^{2}(\Omega_{m},\beta)$ is 310.2
(${\rm dof}=307$). 
In this cosmological scenario the current age of the universe
is found to be $t_{0}\simeq 14.1$Gyr and  
the inflection point is located at $(a_{I},t_{I})\simeq (0.49,0.44t_{0})$.
In Figs. 1b, 1c and 1d we present, for various observational pairs, 
the corresponding likelihood contours (see also Table 1). 

\begin{table*}
\caption[]{Results of the likelihood function analysis. The $1^{st}$ column 
indicates the data used (the last two rows corresponds to the inflection points). 
Errors of the fitted parameters represent $1\sigma$ uncertainties.
Finally, the current age of the universe $t_{0}$ has units of Gyr
(for $H_{0}=70.5$Km/sec/Mpc).}

\tabcolsep 9pt
\begin{tabular}{cccccc} 
\hline
Sample & $\Omega_{m}$& $\beta$ &$t_{0}$ & $a_{I}$& $t_{I}/t_{0}$ \\ \hline \hline 
CMB & uncons. ($\Omega_{m}=0.13$) & $3.58^{+0.04}_{-0.32}$ & 18.2& 0.30 & 0.29 \\
SNIa & $0.20^{+0.12}_{-0.01}$ & uncons. ($\beta=4.6$) & 12.4& 0.59 & 0.50 \\
BAO & $0.28^{+0.11}_{-0.04}$ & uncons. ($\beta=3.50$) & 16.0& 0.36 & 0.33 \\
SNIa-BAO& $0.28^{+0.03}_{-0.02}$ & $3.50^{+0.30}_{-0.36}$ & 14.0& 0.49 & 0.44 \\
CMB-BAO& $0.29^{+0.04}_{-0.03}$ & $3.44^{+0.02}_{-0.02}$ & 14.1& 0.49 & 0.44 \\
SNIa-CMB& $0.29^{+0.01}_{-0.03}$ & $3.44^{+0.02}_{-0.02}$ & 14.1& 0.49 & 0.44 \\
ALL& $0.29^{+0.01}_{-0.02}$ & $3.44^{+0.02}_{-0.02}$ & 14.1& 0.49 & 0.44 \\
\end{tabular}
\end{table*}

Finally, it is worth noting 
that Carneiro et al. (2008) considered 
a different ansatz  
in order to parametrize the time dependence of the vacuum energy. Their
assumption is based on the fact that $\Lambda(t)$ is proportional 
to the Hubble parameter (in our formulation $n_{2}=0$).
They found that this model fits the observational data (BAO+SNIa+CMB) 
at 2$\sigma$ level for $\Omega_{m}\simeq 0.43$. 
In our case, if we marginalize over $\beta=3$ (or $n_{2}=0$), 
then the joint likelihood analysis provides a best fit value of
$\Omega_{m} \simeq 0.35$, but the fit is rather poor 
$\chi_{tot}^{2}(\Omega_{m})\simeq 383$ (${\rm dof}=308$). We investigate 
a bit further this result and we reveal that the poor joint fit is due to 
the fact that the best fit value provided by the 
likelihood analysis of CMB shift parameter 
is found to be more than $3\sigma$ away, $\Omega_{m} \simeq 0.80$ 
(see solid line in figure 2), from the  
SNIa/BAO solution $\Omega_{m} \simeq 0.32$ (see dashed line in figure 2).
This implies that the functional form $E(a)=1-\Omega_{m}+\Omega_{m}a^{-3/2}$ 
fails to fit the CMB data. 
We thus argue that the $\Lambda(t) \propto H(t)$ relation 
produces a discrepancy between the SNIa/BAO and CMB shift parameter 
which may lead to misleading cosmological results.
We further confirm the latter result, 
by using a Bayesian statistics (see for example Davis et al. 2007), in which the 
corresponding estimator is defined as: $BIC=\chi^{2}+k{\rm ln}N$
(where $k$ is the number of parameters and $N$ is the number of 
data points used in the fit). The next step is to estimate the 
relative deviation 
between the two models $\Delta BIC=BIC^{n_{1}H}-BIC^{n_{1}H+n_{2}H^{2}}$.
In general a difference in $BIC$ of $\Delta BIC>6$, is considered 
strong evidence against
that model which ocuurs the larger $BIC$. In our case, 
we find $\Delta BIC \simeq 69$ which implies a strong evidence against 
the $\Lambda(t) \propto H(t)$ model.

\subsection{The standard $\Lambda$-Cosmology}
In this section, we wish to remind the reader
of some basic elements of the concordance $\Lambda$-cosmology
in order to appreciate the differences with the $\Lambda(t)$ cosmology.
In the case of $\Lambda=const$, it is straightforward 
to integrate eq.(\ref{frie344}). 
Therefore, the Hubble function predicted 
by the $\Lambda$ model is
\be
H(t)=\sqrt{\frac{\Lambda}{3}}
\coth\left(\frac{3}{2}\sqrt{\frac{\Lambda}{3}}\;t \right) 
\ee
where $\Lambda=3H^{2}_{0}(1-\Omega_{m})$. 
Then the normalized 
Hubble function is written as 
\begin{eqnarray} 
E_{\Lambda}(a)=\frac{H(a)}{H_{0}}=[1-\Omega_{m}+\Omega_{m}a^{-3}]^{1/2} 
\end{eqnarray} 
while $\Omega_{m}(a)=\Omega_{m}a^{-3}/E^{2}_{\Lambda}(a)$. 
To this end, 
the scale factor of the universe is given by
\begin{eqnarray} 
\label{all} 
a_{\Lambda}(t)=\left(\frac{\Omega_{m}}{1-\Omega_{m}}\right)^{1/3}
\sinh^{2/3}\left(\frac{3H_{0}\sqrt{1-\Omega_{m}}t}{2}\right) 
\end{eqnarray} 
or
\be
t_{\Lambda}(a)=\frac{2}{3\sqrt{1-\Omega_{m}}}H^{-1}_{0}
{\rm ln}\left[\frac{\sqrt{1-\Omega_{m}}+E_{\Lambda}(a)}{a^{-3/2}\sqrt{\Omega_{m}}} \right] \;\;.
\ee

Comparing the $\Lambda$ model with the observational data 
we find that the best fit value is $\Omega_{m}=0.28\pm 0.02$ with 
$\chi_{tot}^{2}(\Omega_{m})\simeq 308.5$ (${\rm dof}=308$) in a good agreement with the 
5 years WMAP data (Komatsu et al. 2009). 
Note, that Davis et al. (2007) using  
the Essence-SNIa+BAO+CMB and a Bayesian statistics found $\Omega_{m}=0.27 \pm 0.04$, while 
Kowalski et al. (2008) utilizing the Union08-SNIa+BAO+CMB obtained
$\Omega_{m}=0.274^{+0.016+0.013}_{-0.016-0.012}$ (for $w\sim -1$). Obviously, our results
coincide within $1\sigma$ errors.

The current age of the universe is given by
\begin{eqnarray} 
t_{0\Lambda}=\frac{2}{3\sqrt{1-\Omega_{m}}}H^{-1}_{0}
{\rm ln}\left(\frac{\sqrt{1-\Omega_{m}}+1}{\sqrt{\Omega_{m}}} \right)
\end{eqnarray} 
while the inflection point takes place at 
\begin{eqnarray} 
      \label{infle}
\;\;\;\;\;\;t_{I\Lambda}=\frac{2}{3\sqrt{1-\Omega_{m}}}H^{-1}_{0}
{\rm ln}\left(\frac{\sqrt{5}+1}{2} \right) \nonumber \\
a_{I\Lambda}=\left[\frac{\Omega_{m}}{2(1-\Omega_{m})}\right]^{1/3} \;\;.
\end{eqnarray}
Therefore, we estimate $t_{0\Lambda}\simeq 13.9$Gyr,
$t_{I\Lambda}\simeq 0.52t_{0\Lambda}$ and $a_{I\Lambda}\simeq 0.58$.
Finally, using the previously described Bayesian statistics we find that
$\Delta BIC=BIC^{n_{1}H+n_{2}H^{2}}-BIC^{\Lambda}\simeq 5$.
This comparison implies a preference for the usual $\Lambda$ cosmology.

\section{Comparison between different types of vacuum}
In this section, we investigate in more detail the 
correspondence of the $\Lambda(t)$ model with the traditional 
$\Lambda$-cosmology (see previous sections) in order to show
the extent to which they compare.

\subsection{Compare the cosmic evolution}
Knowing now the parameter space ($\Omega_{m},\beta$)
we present the evolution of the $\Lambda(t)$  
scale factor seen in the upper panel 
of figure 3 as the solid line. It can been seen that 
it closely resembles the
corresponding scale factor of the $\Lambda$ cosmology (dashed line). 
We have checked the cosmic phases of the 
$\Lambda(t)$ scenario against the concordance cosmology 
by utilizing the 
deceleration parameter, $q(a)=-(1-ad{\rm ln}H/da)$.
The evolution of the deceleration parameter 
is presented in the bottom panel of figure 3, 
while in the insert figure we 
plot the relative deviation of the deceleration parameter, 
$\Delta(q-q_{\Lambda})$, between the two vacuum models.
We find the following phases: (a) at early enough 
times $a<0.21$ the deceleration parameters are both positive with
$q>q_{\Lambda}$, which means that the 
cosmic expansion in the $\Lambda(t)$ model 
is more rapid decelerated than in the $\Lambda$ case,
(b) between $0.21<a<0.49$ the deceleration parameters remain positive
but $q<q_{\Lambda}$, (c) then for $0.49<a<0.58$ the 
traditional $\Lambda$ model remains in the decelerated regime $q_{\Lambda}>0$ 
but the $\Lambda(t)$ is starting to accelerate $q<0$ and (d)
for $0.58<a<0.80$ the deceleration parameters are both negative 
and as long as $q<q_{\Lambda}$
the $\Lambda(t)$ model predicts a
much more acceleration than in the $\Lambda$ model
(the opposite situation seems to hold prior to the present epoch
$0.80<a\le 1$). 
In a special case where $\Delta(q-q_{\Lambda})=0$ [$q=q_{\Lambda}$, either at 
$a\simeq 0.21$ or $a\simeq 0.80$] the two vacuum models predict 
exactly the same expansion of the universe.
From figure 3 it becomes clear that the $\Lambda(t)$ model 
reaches a maximum deviation from the $\Lambda$ cosmology 
prior to $a \sim 0.1$ ($z \sim 9$)
and $a \sim 0.45$ ($z \sim 1.2$). 
Therefore, 
in order to investigate whether the expansion of the observed universe
follows one of the above possibilities, we need a robust extragalactic 
distance indicator at redshifts $z>1.2$.
Finally, the deceleration parameters at the present time are 
$q_{0}\simeq -0.50$ and $q_{0\Lambda}\simeq -0.57$.

\begin{figure}
\mbox{\epsfxsize=8cm \epsffile{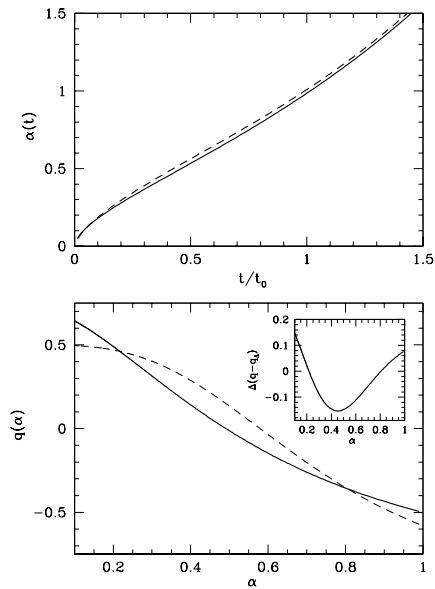}}
\caption{
{\it Upper Panel:} Comparison of the scale factor provided by the 
$\Lambda(t)$ model (solid line) with the 
traditional $\Lambda$ cosmology (dashed line).
{\it Bottom Panel:} The evolution of the deceleration parameter. In 
the insert panel we present the 
relative deviation $\Delta(q-q_{\Lambda})$ of the deceleration 
parameters.}
\end{figure}

\subsection{Compare the linear growth factor}
In the framework of a time varying vacuum,
the corresponding time evolution equation for the mass density contrast, 
in a pressureless fluid is given by (Arcuri \& Waga 1994; see also Borges et al. 2008):
\be\label{eq:11}
\ddot{D}+(2H+Q)\dot{D}-\left[\frac{\rho_{m}}{2}-2HQ-\dot{Q} \right]D=0\;\; .
\ee
where $\rho_{m}=3H^{2}-\Lambda$ (see eq.\ref{frie1}) and 
$Q(t)=-\dot{\Lambda}/\rho_{m}$. It becomes clear, that
the interacting vacuum energy affects the growth factor via 
the function $Q(t)$. Obviously,
in the case of a constant $\Lambda$ [$Q(t)=0$], the above equation reduces 
to the usual time evolution equation for the mass density contrast 
(see Peebles 1993). In this context, 
the growing solution as a function of redshift is given by:
\be\label{eq24}
D_{\Lambda}(z)=\frac{5\Omega_{\rm m} E_{\Lambda}(z)}{2}\int^{\infty}_{z} \frac{(1+x)}{E^{3}_{\Lambda}(x)} 
{\rm d}x\;\;. 
\ee
We now proceed in an attempt to analytically solve eq.(\ref{eq:11}). To do so, 
we change variables from $t$ to a new one following the transformation
\be\label{tran1}
y={\rm exp}(n_{1}t/2) \;\;\;0<y<1 \;\;.
\ee  
Doing so, eq.(\ref{eq:11}) can be written:
\be\label{eq:22}
\beta^{2} y(y-1)^{2}D^{''}+2\beta (y-1)(5y-\beta)D^{'}-2(6-\beta)(\beta-2y)D=0
\ee
where prime denotes derivatives with respect to $y$. 
We find that eq.(\ref{eq:22}) has a decaying 
solution for $\beta<8$ of the form $D_{1}(y)=(y-1)^{(\beta-6)/\beta}$. 
The second independent solution (growing mode) 
of eq.(\ref{eq:22}) can be found easily from the following expression:
\be
\label{eq:ff2} 
D(y)=D_{1}(y)\int_{y}^{1} \frac{(u-1)^{2/\beta}du}{u^{2}} \;\;. 
\ee
\begin{table}
\caption[]{Cosmological data of the growth rate of clustering (see
Nesseris \& Perivolaropoulos 2008).
The correspondence of the columns is as follows: redshift, observed 
growth rate and references.}

\tabcolsep 5pt
\begin{tabular}{ccc} 
\hline
z & $f_{obs}$& Refs.\\ \hline \hline 
0.15 & $0.51\pm 0.11$& Verde et al. 2002; Hawkins et al. 2003\\
0.35 & $0.70\pm 0.18$& Tegmark et al. 2006\\
0.55 & $0.75\pm 0.18$& Ross et al. 2006\\
1.40 & $0.90\pm 0.24$& da Angela 2006\\
3.00 & $1.46\pm 0.29$& McDonald 2005\\
\end{tabular}
\end{table}
Inserting eq.(\ref{frie6}) and eq.(\ref{frie9}) 
into eq.(\ref{tran1}), the $y$ variable 
is related with the scale factor as: 
\be
y=\frac{a^{\beta/2}(1-\Omega_{m})+\Omega_{m}}{\Omega_{m}}\;\;.
\ee
In the redshift regime [$a=(1+z)^{-1}$]
the combination of the above two equations 
lead to the following growing mode:
\be
\label{eqff2} 
D(z)=C(\Omega_{m})(1+z)^{(6-\beta)/2}\int_{z}^{\infty} 
\frac{(x+1)^{(\beta-4)/2}dx}{E^{2}(x)} 
\ee
where 
\be
C(\Omega_{m})=\frac{\beta}{2}\Omega^{2}_{m}\left(\frac{1-\Omega_{m}}
{\Omega_{m}} \right)^{(2\beta-4)/\beta} \;\;\;.
\ee

\begin{figure}
\mbox{\epsfxsize=8cm \epsffile{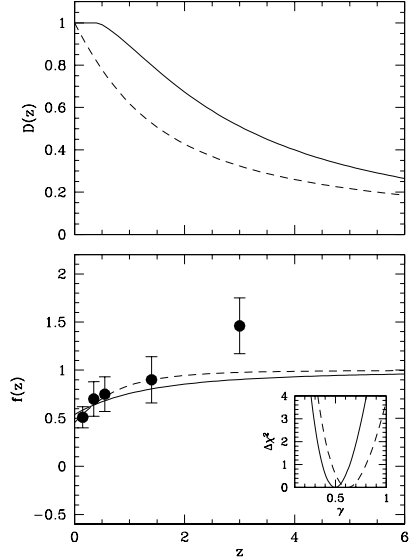}}
\caption{{\it Upper Panel:} 
The evolution of the growth factor 
for different vacuum models. 
The lines correspond to $\Lambda(t)$ (solid) and 
$\Lambda$ (dashed).
{\it Bottom Panel:} 
Comparison of the observed and theoretical evolution of the growth rate of clustering 
$f(z)$. Note, that data can be found in Nesseris \& Perivolaropoulos 
(2008). The best fit for the $\Lambda(t)$ model shows a $\sim 10\%$ 
difference from the traditional $\Lambda$ model, especially at 
large redshifts.}
\end{figure}

In the upper panel of figure 4 we present the growth factor 
evolution, derived by integrating 
eq.(\ref{eq24}) and eq.(\ref{eqff2}), for the two vacuum models. Note that 
the growth factors are normalized to unity 
at the present time. Despite 
the fact that the global cosmological behavior of the $\Lambda(t)$ 
vacuum model is in a good agreement with the usual $\Lambda$ cosmology 
(as it seen in figure 3), the two vacuum cosmological models trace differently 
the evolution of the matter fluctuation field.
In particular, close to the present epoch ($z<0.3$) the 
$\Lambda(t)$ growth factor reaches a plateau, which means that 
the matter fluctuations are effectively frozen. 
It is obvious that the growth factor in the $\Lambda(t)$ model
is much greater than that of the concordance $\Lambda$ cosmology.
Indeed, assuming that clusters of galaxies have formed prior to
the epoch of $z\simeq 1.4$ ($a\sim 0.42$), in which 
the most distant cluster has been found (Mullis et al. 2005; 
Stanford et al. 2006), 
the deviation $(1-D/D_{\Lambda})\%$, of the growth factor
$D(a)$ for the $\Lambda(t)$ scenario 
with respect to the $\Lambda$ solution $D_{\Lambda}(a)$ is
$-51\%$ while prior to the inflection point ($a_{I}\sim 0.5$)
we find $-43\%$. We conclude that the behavior of the growth factor 
is sensitive to the different types of vacuum with
$D(z)>D_{\Lambda}(z)$ and
it is expected that this 
difference will affect also the predictions related with the 
formation of the cosmic structures (see section 5).

\subsection{Compare the growth rate of clustering}
We further compare the two vacuum cosmological scenarios
by utilizing the well known indicator of clustering, 
namely the growth rate $f(a)\equiv d{\rm ln}D/d{\rm ln}a$ (Peebles 1993).
The corresponding parametrization of the growth rate of clustering
can be achieved by introducing a growth index $\gamma$
(see Wang \& Steinhardt 1998) defined by 
\be
f(a)=\Omega^{\gamma}_{m}(a) \;\;.
\ee
In order to quantify the growth index we perform a standard $\chi^{2}$ 
minimization procedure (described previously) between the measured
growth rate of the 2dF and SDSS catalogs 
(see Table 2; Nesseris \& Perivolaropoulos 2008)
with those expected in our spatially flat cosmological models
\be
\chi^{2}(\gamma)=\sum_{i=1}^{5} \left[ \frac{f_{obs}(z_{i})-
f_{\rm model}(z_{i},\gamma)}
{\sigma_{i}}\right]^{2} \;\;,
\ee 
where $\sigma_{i}$ is the observed growth rate uncertainty. 
In the bottom panel of figure 4, we present 
the measured $f_{obs}(z)$ (filled symbols)
with the estimated growth rate 
function $f(z)=\Omega^{\gamma}_{m}(z)$ for 
the considered cosmological models.
Notice, that for the $\Lambda(t)$ 
model (solid line) we use $(\Omega_{m},\beta)=(0.29,3.44)$
and for the $\Lambda$ case (dashed line) we impose $\Omega_{m}=0.28$.
Also, in the insert panel of figure 4 we plot the variation of 
$\Delta \chi^{2}=\chi^{2}(\gamma)-\chi^{2}_{\rm min}(\gamma)$ 
around the best $\gamma$ fit. We find that the growth index 
is $\gamma=0.50^{+0.14}_{-0.12}$ ($\chi^{2}/{\rm dof}=1.14$)
for the $\Lambda(t)$, which is somewhat less (but still 
within $1\sigma$ errors) 
than the $\Lambda$ growth index, $\gamma_{\Lambda}=0.62^{+0.18}_{-0.15}$ 
($\chi^{2}/{\rm dof}=0.75$).

\section{The formation of galaxy clusters}
In this section we attempt to briefly investigate the cluster formation
processes by generalizing the basic equations
which govern the behavior of the matter perturbations
within the framework of a $\Lambda(t)$ flat cosmology.
Also we compare our predictions with those found for the traditional 
$\Lambda$ cosmology. This can help us  
to understand better the theoretical expectations of
the $\Lambda(t)$ cosmological scenario as well as the variants from 
the $\Lambda$ model.

\begin{figure}
\mbox{\epsfxsize=8cm \epsffile{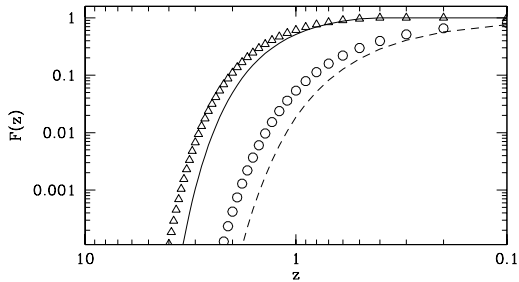}}
\caption{Theoretical predictions of the fractional rate of cluster formation as a 
function of redshift. Note, that the solid and dashed line 
corresponds to $\Lambda(t)$ and usual $\Lambda$ 
cosmological model respectively ($\sigma_{8}=0.80$).
Finally, for $\sigma_{8}=0.95$ we get: (i) the open triangles for $\Lambda(t)$ 
and (ii) open circles for $\Lambda$.}
\end{figure}

The concept of estimating the fractional rate of cluster formation 
has been brought up by different 
authors (cf. Peebles 1984; Weinberg 1987; 
Martel \& Wasserman 1990
Richstone, Loeb \& Turner 1992). The 
above authors introduced a methodology which computes the 
rate at which mass joins virialized structures, which grow from small 
initial perturbations in the universe.
In particular, the basic tool is the so called Press-Schechter formalism 
which considers the fraction of mass in the universe contained
in gravitationally bound structures (such as galaxy clusters) 
with matter fluctuations greater than a critical value $\delta_{c}$. 
Assuming that the density contrast is normally distributed 
with zero mean and variance $\sigma^{2}(\delta)$
we have:
\be\label{eq:88}
dP(\delta)=\frac{1}{\sqrt{2\pi}\sigma} 
{\rm exp}\left[-\frac{\delta_{c}^{2}}{2\sigma^{2}(M,z)} \right]d\delta 
\ee
where $\delta_{c}$ is the linearly extrapolated density 
threshold above which structures collapse, ie, $\delta_{c}=1.686$.
Note, that it has been shown that
$\delta_{c}$ depends only weakly on $\Omega_{m}$ (Eke, Cole \& Frenk 1996).
In this kind of studies it is common to 
parametrize the rms mass fluctuation amplitude at 
8 $h^{-1}$Mpc which can be expressed as a function of redshift as 
$\sigma(M,z)=\sigma_{8}(z)=D(z)\sigma_{8}$. 
The current cosmological models are normalized by 
the analysis of the WMAP 5 years data
$\sigma_{8}=0.80$ (Komatsu et al. 2009).  
The integration of eq.(\ref{eq:88}) provides the fraction of the universe, 
on some specific mass scale, that has already 
collapsed producing cosmic structures (galaxy clusters) 
at redshift $z$ and is given by (see also Richstone et al. 1992):
\be
\label{eq:89}
P(z)=\frac{1}{2} \left[1-{\rm erf} 
\left( \frac{\delta_{c}}{\sqrt{2} \sigma_{8}(z)} \right) \right] \;\;.
\ee
Obviously the above generic of form eq.(\ref{eq:89})
depends on the choice of the background cosmology.
The next step is to normalize the probability to give the number of clusters which
have already collapsed by the epoch $z$ (cumulative distribution), divided 
by the number of clusters which have collapsed at the 
present epoch ($z=0$), $F(z)=P(z)/P(0)$.
In figure 5 we present in a logarithmic scale the behavior of normalized 
cluster formation rate as a function of redshift for the 
two cosmological models. In particular, for the traditional 
$\Lambda$ cosmology we find the known behavior in 
which galaxy clusters appear to be formed at high redshifts $z\sim 2$ (see
for example Basilakos 2003 and references therein), while
the same general picture seems to hold for the $\Lambda(t)$ model.
However, in the latter case we find the following
differences: (i) clusters appear to  
form earlier ($z\sim 3.5$) with respect 
to the $\Lambda$ model 
and (ii) prior to $z\sim 0.4$ the cluster 
formation has terminated 
due to the fact that the matter fluctuation field effectively freezes  
(see section 3.4). It is worth noting that 
the different formation rates between the two vacuum models, is due to
the fact that the evolution of the corresponding growth factors  
are different (see the upper panel of figure 4).
Finally, for a higher $\sigma_{8}$ value ($\sigma_{8}=0.95$) 
the corresponding cluster formation rate 
moves to higher redshifts [see figure 5: $\Lambda(t)$-open 
triangles and $\Lambda$-open points]. 
The opposite situation is true for $\sigma_{8}<0.80$.

\section{Conclusions}
In this paper we study analytically and numerically 
the large and small scale dynamical properties of the 
FLRW flat cosmologies in which the ''vacuum'' energy is a function of 
the cosmic time $\Lambda(t)$. Assuming that the vacuum component 
can be expressed as a power series 
$\Lambda=n_{1}H+(3-\beta)H^{2}$, we find that the 
time evolution of the basic cosmological functions are described in 
terms of exponential functions which  
can accommodate a late time accelerated
expansion, equivalent to the standard $\Lambda$ model.
Performing, a joint likelihood analysis using
the current observational data (SNIa, CMB shift parameter and BAOs), we put
tight constraints on the main cosmological parameters 
of the $\Lambda(t)$ model. In particular, we find 
$\Omega_{m}\simeq 0.29$, $\beta \simeq 3.44$ and the age of the 
universe is $t_{0}\simeq 14.1$Gyr (for $h=H_{0}/100 \simeq 0.705$). 
Also, we compare the $\Lambda(t)$ scenario with the traditional 
$\Lambda$ cosmology. We find that 
the behavior of the global expansion in the $\Lambda(t)$ model 
compares well with that of the usual $\Lambda$ cosmology. However,
there are differences especially when we consider the small scale 
dynamics. Indeed, we reveal that the $\Lambda(t)$ cosmological model 
has two important differences over the considered $\Lambda$ cosmology:

\begin{itemize}

\item The amplitude and the shape of the linear growth 
of perturbations are different with respect to the 
$\Lambda$ solution. As an example, prior to the inflection 
point the $\Lambda(t)$ growth factor increases by a 
factor of $\sim 43 \%$. In this context, the 
growth index of clustering ($\gamma\simeq 0.50$) 
is somewhat different with that of 
the $\Lambda$ model ($\gamma_{\Lambda}\simeq 0.62$).

\item The large scale structures 
(such as galaxy clusters) form earlier ($z\sim 3.5$)
with respect to those 
produced in the framework of the concordance $\Lambda$ model ($z\sim 2$).
\end{itemize}

\section*{Appendix}
In this appendix we treat analytically, as much as possible, the problem of the
time varying $\Lambda(t)$ parameter 
with the aid of a power series in $H$ up to a third order: 
$\Lambda=n_{1}H+n_{2}H^{2}+n_{3}H^{3}$ ($n_{3}\ne 0$). 
The time evolution equation for the Hubble flow 
is obtained by eq.(\ref{frie344}) as:
\be
\label{eq:899}
\int_{+\infty}^{H}\frac{dy}{y(n_{3}y^{2}-\beta y+n_{1})}=\frac{t}{2}
\ee
where $\beta=3-n_{2}$. In particular,
the discriminant $D=\beta^{2}-4n_{1}n_{3}$ characterizes 
the solutions of eq.(\ref{eq:899}) as:

\begin{itemize}
\item {{\bf \it Case 1:} $D>0$ ($\beta^{2}-4n_{1}n_{2}>0$):} The 
corresponding general solution of eq.(\ref{eq:899}) is written
\be 
\label{eq:999}
{\rm ln}\left[\frac{H^{\frac{1}{\rho_{1}\rho_{2}}}
(H-\rho_{1})^{\frac{1}{\rho_{1}(\rho_{1}-\rho_{2})}}}
{(H-\rho_{2})^{\frac{1}{\rho_{2}(\rho_{1}-\rho_{2})}}}\right]=\frac{n_{3}t}{2} 
\ee
where $\rho_{1,2}=\frac{\beta \pm \sqrt{D}}{2n_{3}}\ne 0$. 
As an example for $\beta=0$ (or $n_{2}=3$, $\rho_{1}=-\rho_{2}$) 
we obtain
\be
H(t)=\frac{\rho_{2}}{\sqrt{1-{\rm e}^{n_{3}\rho^{2}_{2}t}}}
\ee
and 
\be
a(t)=a_{1}\left(\frac{1+\sqrt{1-{\rm e}^{n_{3}\rho^{2}_{2}t}}}
{1-\sqrt{1-{\rm e}^{n_{3}\rho^{2}_{2}t}}} \right)^{-\frac{1}{n_{3}\rho_{2}}} \;\;,
\ee
where $n_{3}<0$ and $\rho_{2}>0$.
We would like to point out that 
as long as the cosmic time takes large values ($t\gg 1$),
the $\Lambda(t)$ model has the de-Sitter
feature due to $a(t) \sim {\rm e}^{\rho_{2}t}$. On the other
hand, it is very interesting the fact that this model 
has no initial singularity. Indeed, 
for $t \longrightarrow 0$ we get $a(t) \longrightarrow a_{1}$. 

Now, if we consider $\beta \ne 0$ then 
the situation becomes complicated (see eq.\ref{eq:999}) but for the special case of 
$\rho_{1}=2\rho_{2}$ we can derive the following analytical solutions: 
\be
H(t)=\rho_{2}+\frac{\rho_{2}}{\sqrt{1-{\rm e}^{n_{3}\rho^{2}_{2}t}}}
\ee
and 
\be
a(t)=a_{1}{\rm e}^{\rho_{2}t}
\left(\frac{1+\sqrt{1-{\rm e}^{n_{3}\rho^{2}_{2}t}}}
{1-\sqrt{1-{\rm e}^{n_{3}\rho^{2}_{2}t}}} \right)^{-\frac{1}{n_{3}\rho_{2}}}
\ee
where $n_{3}<0$, $\rho_{2}>0$ and $\beta <0$.
Again, the $\Lambda(t)$ model asymptotically reaches the de-Sitter regime
$a(t) \sim {\rm e}^{2\rho_{2}t}$, while 
for $t \longrightarrow 0$ we again find no singularity
$a(t) \longrightarrow a_{1}$. 

\item {{\bf \it Case 2:} $D=0$ ($\beta^{2}=4n_{1}n_{3}$):} In this 
case the integration of eq.(\ref{eq:899}) leads to the solution of:
\be
{\rm ln}\left(\frac{H}{H-\rho}\right)-\frac{\rho}{H-\rho}=\frac{\rho^{2}n_{3}t}{2}
\ee
where $\rho=\frac{\beta}{2n_{3}}\ne 0$. Now if $\beta=0$ ($\rho=0$), 
which implies that $n_{1}=0$, then the solution of eq.(\ref{eq:899}) 
is given by
\be
H(t)=\sqrt{-\frac{1}{n_{3}t}}   \;\;\;\;\;,\;\;n_{3}<0
\ee
and 
\be
a(t)=a_{1}{\rm e}^{\sqrt{-\frac{4}{n_{3}}t } } \;\;.
\ee

\item {{\bf \it Case 3:} $D<0$ ($\beta^{2}-4n_{1}n_{2}<0$):} 
In this case the integration of 
eq.(\ref{eq:899}) leads to the solution of:
\be 
{\rm ln}\left( \frac{n_{3}H^{2}}{n_{3}H^{2}-\beta H+n_{1}} \right)
+\frac{2\beta}{\sqrt{-D}} \left[ {\rm tan}^{-1}G(H)-\frac{\pi}{2}
\right]=n_{1}t
\ee
where $G(H)=(2n_{3}H-\beta)/\sqrt{-D}$.

\section*{Acknowledgements} 
I thank the referee for his/her very detailed report, useful comments and
suggestions.

\end{itemize}

\end{document}